\documentclass{aa}

\usepackage{psfig}
\usepackage{graphicx}
\usepackage{natbib}
\usepackage[english]{babel}  
\usepackage{times}

\begin{document}

\title{Generation of radiative knots in a randomly pulsed protostellar
jet}

\subtitle{II. X-ray emission}

\author{R. Bonito\inst{1, 2} \and S. Orlando\inst{2} \and M. Miceli\inst{1, 2}
\and J. Eisl\"offel\inst{3} \and G. Peres\inst{1, 2} \and F. Favata\inst{4}}

\offprints{R. Bonito\\ \email{sbonito@astropa.unipa.it}}

\institute{Dip. Scienze Fisiche ed Astronomiche, Sez. Astronomia,
Universit\`a di Palermo, P.zza del Parlamento 1, 90134
Palermo, Italy
\and 
INAF -- Osservatorio Astronomico di Palermo, P.zza del Parlamento 1,
90134 Palermo, Italy 
\and
Th\"uringer Landessternwarte, Sternwarte 5, D-07778 Tautenburg, Germany
\and
European Space Agency
Community Coordination and Planning Office
8-10 rue Mario Nikis
F-75738 Paris cedex 15
France
} 

\date{Received, accepted}

\authorrunning{}
\titlerunning{}

\abstract
{Protostellar jets are known to emit in a wide range
of bands, from radio to IR to optical bands, and to date also about ten X-ray emitting jets have
been detected, with a rate of discovery of about one per year.}
{We aim at investigating the mechanism leading to the
X-ray emission detected in protostellar jets and, in particular,
at constraining the physical parameters that describe the
jet/ambient interaction by comparing our model predictions with
observations available in the literature.}
{We perform 2D axisymmetric hydrodynamic simulations of the
interaction between a supersonic jet and the ambient medium. The jet
is described as a train of plasma blobs randomly ejected by the
stellar source along the jet axis. We explore the parameter space
by varying the ejection rate, the initial Mach number of the
jet, and the initial density contrast between the ambient medium
and the jet.
We synthesized from the model the X-ray emission as it would be observed with the current X-ray telescopes.}
{The mutual interactions among the ejected blobs and of the blobs
with the ambient medium lead to complex X-ray emitting structures within
the jet. The X-ray sources consist of several components: irregular
chains of knots; isolated knots with measurable proper
motion; apparently stationary knots; reverse shocks. The predicted X-ray
luminosity strongly depends on the ejection rate and on the initial
density contrast between the ambient medium and the jet, with a
weaker dependence on the jet Mach number.}
{Our model represents the first attempt to describe the
X-ray properties of all the X-ray emitting protostellar jets discovered
so far. The comparison between our model predictions and the observations
can provide a useful diagnostic tool necessary for a proper interpretation
of the observations. In particular, we suggest that the
observable quantities derived from the spectral analysis of X-ray
observations can be used to constrain the ejection rate, a parameter
explored in our model that is not measurable by
current observations in all wavelength bands.}


\maketitle

\section{Introduction}
\label{Introduction}

In the past decade, high-energy emission from protostellar jets
(originally suggested by \citealt{pm81}) has been discovered, taking
advantage of both the high spatial resolution of Chandra and the high
effective area of XMM-Newton. To date, about ten X-ray emitting
Herbig-Haro (HH) objects are known and the main properties of their
emission are reviewed in Table \ref{tab-HH-X} (see also \citealt{bop07}).

\begin{table*}[!t]
\begin{center}
\caption{Physical properties observed in X-ray emitting HH jets.}
\label{tab-HH-X}
\par
\begin{tabular}{lccccccccl}
\hline
\hline
object   &  $t_{\rm exp}$ & $cnts$       & $L_{\rm X} $            & $T_{X}$   & $L_{\rm j}$ & $z_{\rm X}$ & $D$   & $LM/HM$ & Reference \\
         &  [ks]          &              &[$10^{29}$ erg s$^{-1}$] & [$10^{6}$ K] & [arcsec]    & [AU]        &  [pc]  &         \\
\hline
HH 2     & $21.4$         & $11$         & $5.2$   	           & $1$	   & $2$     & $56000^c$   & $480$  &  LM & \citet{pfg01} \\
HH 154   & $97$           & $63$         & $3$	     	           & $4$	   & $5$     & $70-140$    & $140$  &  LM & \citet{fbm06}$^d$ \\
HH 80/81 & $37.3$         & $46/63$      & $450/430$	           & $1.5$	   & $-$     & $515000$    & $1700$ &  HM & \citet{ptm04} \\
HH 168   & $78$           & $-$          & $30$		           & $6.5$	   & $-$     & $-$	   & $730$  &  HM & \citet{ptu09}  \\
HH 210   & $838$          & $31$         & $10$	                   & $0.8$         & $-$     & $37000^c$   & $450$  &  HM & \citet{gfg06} \\
HH 540   & $795.8$        & $\approx200$ & $0.4$                   & $6.6$	   & $-$     & $<90$       & $450$  &  LM & \citet{kfg05}$^a$ \\
HH 216   & $78$           & $8$          & $10$    	           & $-$	   & $4$     & $-$	   & $2000$ &  -  & \citet{lgm07} \\
DG Tau   & $90$           & $18; 9$$^b$  & $0.12$   	           & $3.4$	   & $5$     & $30$        & $140$  &  LM & \citet{gsa08} \\
Z CMa    & $39.6$         & $20$         & $>2.5$   	           & $2.3$	   & $-$     & $> 2000$    & $1050$ &  HM & \citet{sho09} \\
TKH 8    & $89.2$         & $28$         & $20$   	           & $35.4$        & $-$     & $450-900$   & $450$  &  LM & \citet{tkk04}$^e$ \\
\hline
\end{tabular}
\end{center}
\smallskip 
$t_{\rm exp}$ is the exposure time of the observations (or the total exposure
time adding together several observations, as in the case of DG Tau);
$cnts$ are the collected photons; $L_{\rm X}$ is the X-ray luminosity
in the $[0.3-10]$ keV band; $T_{X}$ is the best fit
temperature; $L_{\rm j}$ is the linear size of the X-ray source; $z_{\rm
X}$ is the distance of the X-ray source from the central protostar;
$D$ is the distance of the HH object; $LM/HM$ indicates the low-mass or high-mass
young stellar object from which the jet originates.

$^a$ Improved estimates of $L_{\rm X}$ and $T_{X}$ from Joel Kastner and Ettore Flaccomio, private communication
$^b$ The first value is associated with the South-Western jet and the second with the North-Eastern jet in DG Tau
$^c$ Values derived from SIMBAD
$^d$ See also \citet{bfr03}
$^e$ See also \citet{tkh01}

\end{table*}

The X-ray sources detected in HH jets are characterized
by different morphologies, luminosities, and locations within the jet. In some cases, the X-ray emitting region is located at
the base of the (optical) jet, near the protostar from which the jet
originates. This is the case, for instance, for the low-mass
(LM) HH 154 and DG Tau jets, both in Taurus (see \citealt{bfr03,
gsb05}). In other cases, the X-ray source is located further
away from the protostar as, for example, for the high-mass (HM) HH 80/81
jet. As for the morphology, some of the X-ray sources
cannot be resolved by current instruments (and appear point-like),
some others appear to be elongated.  In one case (HH 154)
a knotty X-ray source is found (resembling the knotty morphology
commonly observed in the optical emission from HH jets) which
consists of an elongated tail with a measurable proper motion away
from the stellar source and an apparently stationary point-like source
(\citealt{fbm06}).

Current models, aimed at explaining the origin of X-ray emission
from protostellar jets, considered a continuous supersonic jet
propagating through a homogeneous interstellar medium (\citealt{bop04,
bop07}). These models succeeded in predicting X-ray sources associated
with jets with luminosity and proper motion in good agreement with the
observations. They failed, however, in reproducing the
complex knotty morphology of the X-ray source detected
in HH 154 (\citealt{fbm06}).

In a previous paper (\citealt{bop10}; hereafter Paper I), we proposed
an improvement to these models by considering the scenario based on
a pulsed jet, i.e. a jet characterized by an ejection velocity
varying randomly in time, and interacting with an initially homogeneous
ambient medium. The aim was to investigate the origin of the irregular
knotty structure observed in protostellar jets in different wavelength
bands and the complex interactions occurring among blobs of plasma ejected
from the stellar source. Our analysis showed that the mutual interactions
of blobs ejected at different epochs and with different speed naturally reproduce the irregular pattern of knots observed along the jet axis in
many HH objects and lead to a variety of plasma components which cannot be
described by models of a jet ejected with a sinusoidal variable velocity
(e.g. \citealt{rdk07}).

Here we further explore the model of a randomly ejected pulsed
jet presented in Paper I with the aim to investigate the properties
of its X-ray emission and how these properties may depend on the
physical conditions in which the jet evolves. In fact, X-ray emission from
jets is observed in HH objects at different evolutionary stages and with
different masses: for instance, HH 154 originates from a LM binary system
of class 0 and class I sources; DG Tau is a more evolved LM classical
T Tauri star; HH 80/81 originates from a HM protostar. Given the wide
range of conditions in which X-ray emission from jets originates, it is
important to address the following issues: where and when is the X-ray
emission expected to arise from HH jets? How common is the high-energy
emission from HH jets?

The paper is organized as follows: in Sect. \ref{The model} we present
our model; in Sect. \ref{Results} we show our results concerning the
morphology and luminosity in the X-ray band; in Sect. \ref{Discussion}
we discuss our results and in Sect. \ref{Conclusions} we draw our
conclusions.

\section{The model}
\label{The model}

We model the evolution of a protostellar jet ejected with a
supersonic random speed and ramming into an initially homogeneous ambient
medium. A detailed description of our model can be found in Paper I
to which the reader is referred for more details. Briefly, we
performed 2D hydrodynamic simulations of the interaction between the
jet and the ambient medium in a cylindrical coordinate system, $(r,
z)$, by solving the equations of conservations of mass, momentum, and
energy, using the FLASH code (\citealt{for00}). The model takes
into account the radiative losses from optically thin plasma
and the thermal conduction in both the Spitzer (\citealt{spi62})
and saturated (\citealt{cm77}) regimes.

Axisymmetric boundary conditions are imposed along the jet axis
(consistent with the adopted symmetry), inflow boundary conditions at the
base for $r < r_{j}$, where $r$ is the radial distance from the jet axis
in cylindrical coordinates and $r_{j} \approx 30$ AU is the jet radius,
and outflow boundary conditions elsewhere (see also Paper I). The spatial
resolution achieved in our simulations is about $8$ AU, i.e. about $8$
times the resolution of Chandra/ACIS-I observations at a distance
of $\approx150$ pc (i.e. roughly the distance of the closest star-forming region; SFR).

The time covered by our simulations ranges between $100$
yr and $400$ yr, depending on the initial setup of the
model. The jet is described as a train of blobs, each lasting for
$0.5$ yr, with an ejection rate corresponding to a time interval
between the ejection of two consecutive blobs $\Delta t = 0.5,
2, 8$ yr. The initial setup is derived from the analysis
of the optical and X-ray data and from the results of previous models
(\citealt{bop07}). Each blob is ejected with a random velocity
directed along the jet axis (the $z$-axis), sampled from an
exponential distribution, and with a maximum velocity $v_{j}$,
corresponding to the initial Mach number $M = v_{j}/c_{s}$, where
$c_{s}$ is the isothermal sound speed (see Paper I for more
details). 

We explore both the initial light jet scenario (a jet
initially less dense than the unperturbed ambient medium; hereafter LJ
runs), and the initial heavy jet scenario (a jet initially denser than
the ambient medium; hereafter HJ runs). Note that the concept of light and
heavy jet is referred to the initial conditions of our simulations. In
fact, as already discussed in Paper I, the density contrast between the
ambient medium and the blob can vary during the jet/ambient evolution,
the density of the medium varying by several orders of magnitude after
the first high-speed blob has perturbed the whole computational domain
(see Fig. 3 of Paper I). In other words, the ejected blobs can be denser or less dense than the medium in which they propagate, in
each of the simulations considered here. Table \ref{tab_mod} summarizes
the physical parameters characterizing the simulations. In all
the models we assume: $r_{\rm j}\approx30$ AU as initial jet radius and we
derive $T_{\rm a} = 10^{3}$ K as initial ambient temperature, assuming
initial pressure balance between the jet and the ambient medium.

\begin{table}[!t]
\caption{Summary of the initial physical parameters characterizing the simulations (see text for details).}
\label{tab_mod}
\begin{center}
\begin{tabular}{lccccccc}
\hline
\hline
Model       & $\Delta t$ & $\nu$  & $M$    & $v_{\rm j}$   & $T_{\rm j}$ & $n_{\rm a}$  \\
            & [yr]         &        &        & [km s$^{-1}$] & [K]  & [cm$^{-3}$] \\
\hline
LJ0.5-M1000 & $0.5$      & $10 $  & $1000$ & 10 - 4680   & $10^{4}$    & $5000$ \\
LJ2-M1000   & $2  $      & $10 $  & $1000$ & 10 - 4680   & $10^{4}$    & $5000$ \\
LJ8-M1000   & $8  $      & $10 $  & $1000$ & 10 - 4680   & $10^{4}$    & $5000$ \\	  
LJ2-M100    & $2  $      & $10 $  & $100 $ & 10 -  470   & $10^{4}$    & $5000$ \\
LJ2-M300    & $2  $      & $10 $  & $300 $ & 10 - 1400   & $10^{4}$    & $5000$ \\
LJ2-M500    & $2  $      & $10 $  & $500 $ & 10 - 2340   & $10^{4}$    & $5000$ \\
HJ2-M300    & $2  $      & $0.1$  & $300 $ & 10 - 1400   & $10^{2}$    & $50  $ \\
HJ2-M500    & $2  $      & $0.1$  & $500 $ & 10 - 2340   & $10^{2}$	 & $50  $ \\
HJ2-M1000   & $2  $      & $0.1$  & $1000$ & 10 - 4680   & $10^{2}$ 	 & $50  $ \\
\hline
\end{tabular}
\end{center}
\smallskip
$\Delta t$ is the time interval between two consecutive blobs, $\nu$ is the
ambient-to-jet density contrast, $M$ is the Mach number of the first blob,
$v_{\rm j}$ is the velocity of each ejected blob (here we indicate the
range of values randomly generated by our model), $T_{\rm j}$ is the
initial jet temperature, and $n_{\rm a}$ is the ambient density.
\end{table}

\section{Results}
\label{Results}

The dynamics and energetics of the randomly ejected pulsed jet have
been extensively described in Paper I. Here we focus on the analysis
of the X-ray emission predicted to arise from the pulsed jet. The
X-ray emission is synthesized from the model results by adopting the
methodology described in \citet{bop07} that allows us to derive images and spectra of the
X-ray sources associated with the jets. Briefly, we first recover
the 3D spatial distributions of density and temperature by rotating the
corresponding 2D distributions around the symmetry $z$ axis ($r=0$).
Then we integrate along the line of sight\footnote{Assumed to be
perpendicular to the jet axis}, under the assumption of an optically
thin plasma, and derive the emission measure distribution as a function
of the temperature, EM$(T)$\footnote{See \citet{bop07} for definitions
and details on the EM.}. Using available spectral emission
codes and taking into account both the instrumental response and the
interstellar absorption, we derive the X-ray spectra, the morphology,
and the evolution of the X-ray luminosity, $L_{\rm X}$, to be
compared directly to the parameters derived from the observations.

\subsection{Morphology of X-ray jets}
\label{Spatial distribution of X-ray emission}

\begin{figure}[!t]
\centerline{\psfig{figure=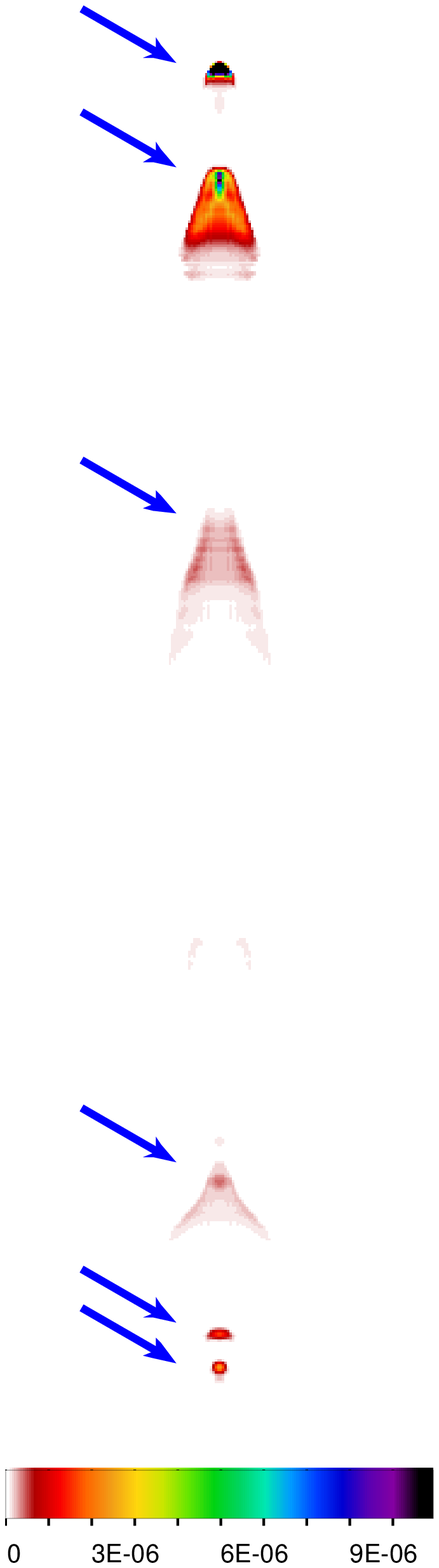,width=2.9cm}
            \psfig{figure=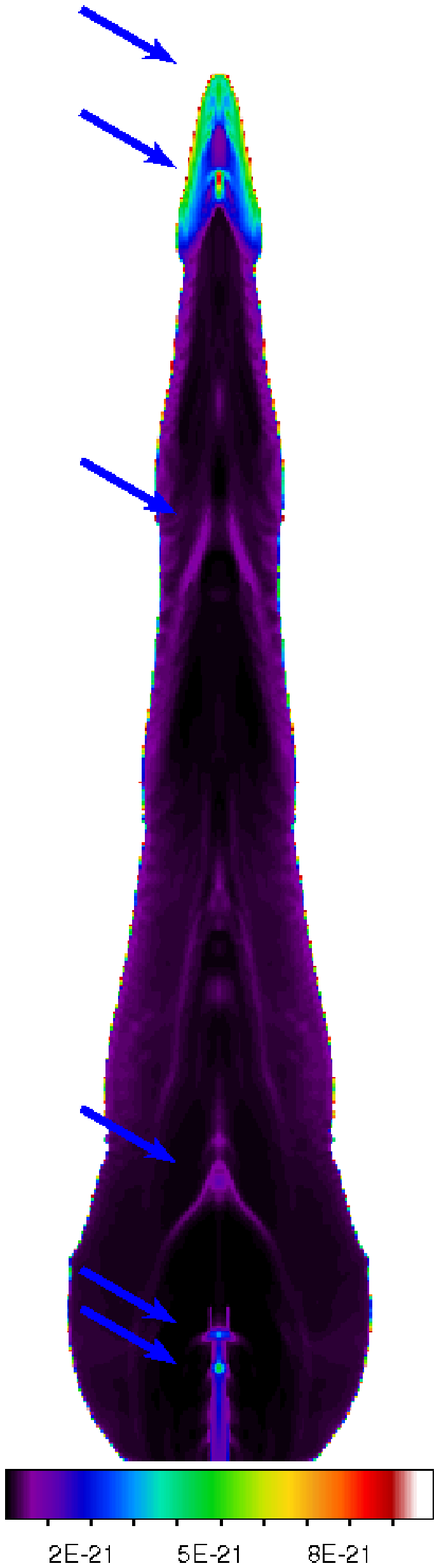,width=2.9cm}}
\caption{X-ray image of the jet synthesized from the pulsed jet model in the [$0.3-10$] keV band (left panel) and
corresponding mass density distribution (right panel) $31$ yr after
the beginning of the simulation in run LJ0.5-M1000. The blue
arrows superimposed on each panel mark the same positions
in the X-ray map and mass density distribution.
The size of the jet is about $4600$ AU, corresponding to $\approx30''$ at $150$ pc.}
\label{X-05-31}
\end{figure}

\begin{figure*}[!t]
\centerline{\psfig{figure=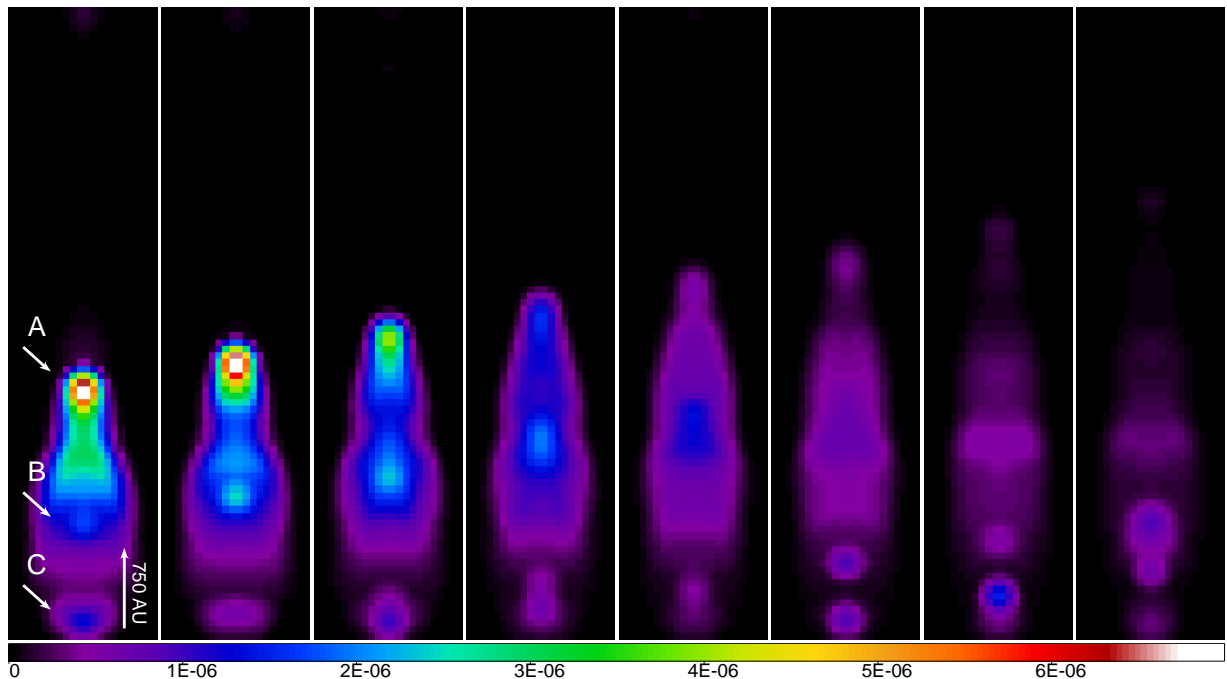,width=16cm}}
\caption{Evolution of the X-ray emission as it would be observed
with Chandra/ACIS-I in run LJ0.5-M1000. 
The head of the jet is outside the spatial domain. Two consecutive panels are
separated by a time interval of $0.5$ yr, the first panel corresponding
to $44$ yr since the beginning of the jet/ambient interaction. 
The spatial scales are shown in the first frame as well as the three main sources highlighted by the arrows: knot A, B, and C.}
\label{X-05}
\end{figure*}

We investigate the morphology of the X-ray sources associated
with jets, by simulating observations with Chandra/ACIS-I,
namely the current X-ray instrument with the highest spatial
resolution. As an example, Fig. \ref{X-05-31} shows the X-ray emission in
the [$0.3-10$] keV band (left panel) and the corresponding mass density distribution (right panel) for run LJ0.5-M1000, assuming a distance of $150$ pc and an absorption
column density $N_{H} = 1.4\times10^{22}$ cm$^{-2}$, as derived for the HH 154 jet. In general the morphology of the source is quite complex
showing several X-ray emitting plasma components: knots resulting
from the interaction among different blobs\footnote{As discussed
in Paper I, we use the term "blob" to indicate the ejected clump
of plasma and "knot" to indicate the observable structure formed along
the jet.} of plasma ejected from the stellar source (for instance see left panel in Fig. \ref{X-05-31}), reverse shocks
interacting with outgoing knots, stationary knots, oblique structures.

The most striking feature in Fig.~\ref{X-05-31} (left panel) is
the irregular pattern of knots aligned along the jet axis and possibly
interacting with each other, analogous to the well-known optical knotty
structure observed within HH jets. Such a complex chain of X-ray knots
originates from the interaction of different ejected blobs with each
other and with the inhomogeneous medium in which they propagate. In fact,
as explained in Paper I, the initially homogeneous ambient medium becomes
quickly inhomogeneous because the blobs ejected with different speeds
and at different epochs perturb the medium and strong variations of the
pre-shock conditions occur (see also Fig. 3 in Paper I). Chains of knots
are particularly evident in cases of high ejection rates; for instance,
we found up to six knots for the case with $\Delta t = 0.5$ yr (see left
panel in Fig. \ref{X-05-31}). The X-ray sources along the jet axis are
associated with strong shocks visible in the density distribution (see
right panel of Fig. \ref{X-05-31}). In general, the brightest
X-ray knots occur at the base of the jet and originate from the ejection of
single blobs of plasma with high speed into the ambient medium. X-ray
emission from isolated knots created by a single high-speed ejected blob
is common to all the runs discussed here.

In principle, it is possible to derive the proper motion of modeled
X-ray knots and compare it to the measured or expected proper motion from
observed X-ray sources in protostellar jets. Our simulations have
shown that the knot speed ranges between $300$ and $3000$ km s$^{-1}$ (i.e. between $0.4$''/yr and $4$''/yr at $150$ pc). As
an example, Fig. \ref{X-05} shows the evolution of the X-ray emission
for run LJ0.5-M1000; the proper motion of X-ray emitting knots A, B, and C is
measurable in a few years. The knots
speed deduced from these measurements ranges between about $300$
km s$^{-1}$ for the C knot at the base of the jet, and $2000$ km s$^{-1}$ (i.e. $3$''/yr at $150$ pc) for the bright A knot. 
It is worth noting that the faint and slow C knot at the base of the jet is persistent throughout the evolution of the jet shown in Fig. \ref{X-05}, due to continuous fueling by subsequent ejected blobs; on the other hand the brightest source (A knot), which shows the highest proper motion, cools down and its X-ray luminosity drops down by about an order of magnitude after about two years.
Unfortunately, up to now, the only X-ray emitting jet for
which it was possible to measure the proper motion of the X-ray source
($\approx 500$ km s$^{-1}$) is HH 154 (\citealt{fbm06}). For this
case, our model predictions are in good agreement with the observations
being the characteristic values of the average knot speeds derived from our model consistent with a few hundreds km/s (see Fig. 10 of Paper I). Nonetheless, our analysis
suggests that, in general, a measurable proper motion of X-ray knots is
expected and our predictions can be challenged in future observations
(see, however, possible misinterpretations of observations
discussed in Sect.~\ref{Main components of X-ray emission of the pulsed
model}).

\begin{figure}[!t]
\centerline{\psfig{figure=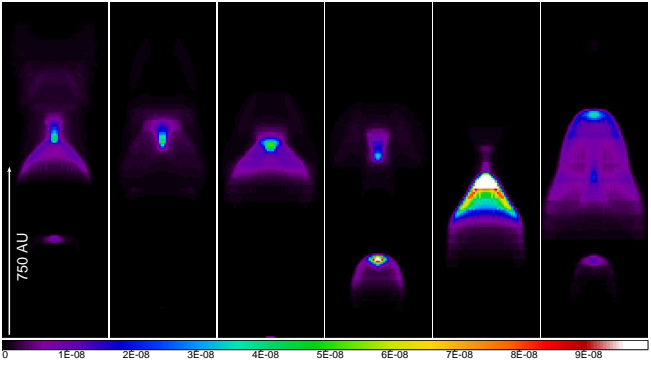,width=8.5cm}}
\caption{Zoom of the X-ray emission at the base of the jet for
run LJ2-M1000 over a time baseline of about 5 yr.
The spatial scale is shown in the first frame ($750$ AU correspond to $5''$ at $150$ pc).}
\label{X-2-zoom}
\end{figure}

Additional contributions to X-ray emission may come from a variety
of complex plasma structures formed within the jet due to interactions
among supersonic blobs, shocks, and the cocoon enveloping the jet. In
particular, oblique structures can form at the cocoon (see also Paper I),
as well as reverse shocks traveling in the opposite direction of
ejected blobs (i.e. toward the stellar source) and possibly interacting
with them. Under certain circumstances, shocks formed within the jet
may show no proper motion on time-scales of a few years. As an example,
Fig. \ref{X-2-zoom} shows in run LJ2-M1000 an almost stationary
X-ray source over a time-scale of about $5$ yr due to the interaction of
a reverse shock with outgoing plasma blobs.

\begin{figure}[!t]
\centerline{\psfig{figure=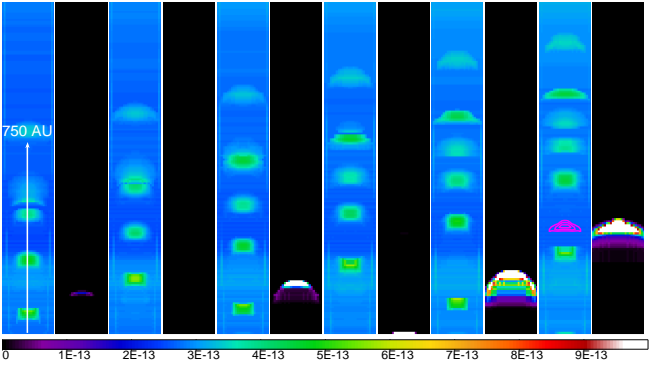,width=8.5cm}}
\caption{Evolution of both expected optical knots (odd panels) and
X-ray emission (even panels) in run HJ2-M300.
The contours of the brightest regions of the X-ray source in the last panel are superimposed on the corresponding optical image. 
The spatial scale is shown in the first panel.}
\label{X-ott}
\end{figure}

The position of the X-ray sources can be compared to that of expected
optical knots, by deriving density maps of plasma with temperature ranging
between $(5\times 10^3 - 10^5)$ K, which is a proxy of optical emission
(see Paper I). As an example, Fig.~\ref{X-ott} shows the evolution of both
expected optical knots and X-ray sources in run HJ2-M300. In general,
the X-ray sources are not co-spatial with optical knots. In fact,
the figure shows a weak X-ray source located at the base of the chain
of optical knots, showing a stratification of the X-ray and
optical emission.

\subsection{X-ray light curves}
\label{LX}

We derived the X-ray luminosity, $L_{X}$, and its evolution for all
the runs listed in Table ~\ref{tab_mod}. Our model predicts significant
X-ray emission in the light jet scenario (LJ runs in Table~\ref{tab_mod})
and very faint or no X-ray emission in the heavy jet scenario (HJ runs in
Table~\ref{tab_mod}; compare the different scales in Fig. \ref{X-05-31}
and Fig. \ref{X-ott}). In this section, we therefore focus on LJ
runs that predict X-ray emission detectable with current X-ray
observatories. 

\begin{figure}[!t]
\centerline{\psfig{figure=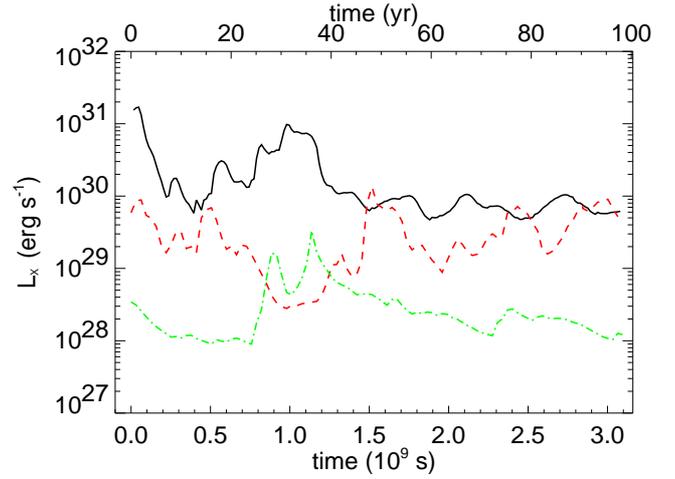,width=9cm}}
\caption{Evolution of the X-ray luminosity derived from runs LJ0.5-M1000
(solid line in black), LJ2-M1000 (dashed line in red), and LJ8-M1000 (dashed-dotted line in green).
Time $= 0$ corresponds to $2.5$ yr, $10$ yr, and $40$ yr from the beginning of the simulation for run LJ0.5-M1000, LJ2-M1000, and LJ8-M1000, respectively.}
\label{Lvst-3-intert} 
\end{figure}

In our simulations, the maximum value of $L_{X}$ is reached
when the first blob (characterized by the highest allowed velocity)
is ejected in the unperturbed ambient medium. Since such a high-speed
blob and its interactions with subsequent ones cause the formation
of very bright sources that are clearly due to the initial transient
configuration, we removed the contribution to $L_{X}$ due to the first
three blobs ejected\footnote{For this reason the epoch time $= 0$ in Fig. \ref{Lvst-3-intert} corresponds to $2.5$ yr, $10$ yr, and $40$ yr from the beginning of the simulation for run
LJ0.5-M1000, LJ2-M1000, and LJ8-M1000, respectively}. Figure \ref{Lvst-3-intert}
shows $L_{X}$ as a function of time for runs LJ0.5-M1000 (solid line in black),
LJ2-M1000 (dashed line in red), and LJ8-M1000 (dashed-dotted line in green), after the transient phase has
been removed. Typical values of $L_{X}$ range between $10^{28}$
erg $s^{-1}$ and a few $10^{31}$ erg $s^{-1}$, in good agreement with
almost all the observations of X-ray emitting HH jets.

Figure~\ref{Lvst-3-intert} shows that $L_{X}$ depends on the velocity
of the ejected blobs and on the ejection rate $\Delta t$. In particular,
the X-ray luminosity decreases by three orders of magnitude from $\Delta
t = 0.5$ yr to $\Delta t = 8$ yr, whereas it can vary by a factor
$10$ in each simulation due to the velocity variations of the ejected
blobs. This small variation of the X-ray luminosity with the velocity is also found by comparing
the values of $L_{X}$ derived from simulations with the same ejection
rate ($\Delta t = 2$) and different Mach numbers, $M = 100, 300, 500$
(runs LJ2-M100, LJ2-M300, and LJ2-M500 in Table~\ref{tab_mod}). We
conclude, therefore, that the critical parameter in determining the
X-ray luminosity of a protostellar jet is the ejection rate of plasma
blobs rather than their velocities: the higher the ejection rate, the
brighter the X-ray source associated with the jet. In fact,
the X-ray emission originates mainly from multiple interactions among
knots and blobs which increase for higher ejection rates.

\section{Discussion}
\label{Discussion}

\subsection{Visibility of X-ray emission from HH jets}
\label{Visibility of X-ray emission from HH jets}

Although hundreds of HH jets have been revealed in several bands\footnote{A catalogue of HH objects has been prepared by B. Reipurth
and it can be found in http://casa.colorado.edu/hhcat/}
(from radio to IR, to optical wavelength bands) up to
now, only ten HH
objects are known to emit also in X-rays (Table \ref{tab-HH-X};
see also \citealt{bop07}). The small fraction of jets visible
in the X-ray band poses a problem because, in principle, all the
high speed jets should emit X-rays. Of course, it is possible
that X-ray emission from HH jets can be observed only under favorite
conditions. In fact, X-ray emitting jets have to be sufficiently
luminous in order to be detected in far away SFRs. Chandra observations of HH 154, the most
luminous X-ray emitting HH jet detected in the nearest SFR, collected about $60$
counts in $100$ ks (\citealt{fbm06}); an even worse statistics have been
obtained for the less luminous X-ray jet revealed in the same SFR and
associated with DG Tau (\citealt{gsa08}). These sources, therefore, are intrinsically faint and analogous ones located at larger distances cannot be detected with current
instruments. It is therefore not surprising that X-ray jets located
further away are very bright, as in the case of the most luminous X-ray
emitting HH object revealed up to now, HH 80/81 (\citealt{ptm04}).

Our simulations have shown that the necessary conditions to have
detectable X-ray emission from a randomly pulsed jet are both a
high ejection rate and a high density of the
medium in which subsequent plasma blobs are ejected. As an example, in a light jet (i.e. the scenario leading to detectable X-ray emission) the X-ray emission produced by a single blob interacting with the surrounding medium can be rather faint, or not present
at all, if the ambient-to-blob density contrast is low, unless the blob propagates with very high speeds ($> 2000$
km s$^{-1}$). 

As for the optical emission, the train of
blobs forming the jet builds up a cocoon enveloping the jet and the
subsequently ejected blobs. In the light jet scenario, such a cocoon
is characterized by high mass density and is the dominant component
in the optical emission, thus making negligible in this band
any contribution from internal knots. Conversely, in the heavy
jet scenario, the density of the cocoon is much lower than that of the
ejected blobs and the main contribution to optical emission comes from
the dense knots formed within the jet (see Fig. \ref{X-ott}). Our
model therefore predicts that light jets can lead to X-ray emission to
observed levels, but do not produce observable optical knots,
whereas heavy jets reproduce the chains of optical knots commonly observed
in almost all the known protostellar jets, but do not lead to
detectable X-ray emission.

The different characteristics of light and heavy jets, described
above, originates from the fact that in the former ones the traveling knots
are decelerated by a factor of about three by the surrounding dense
medium, whereas in the latter ones the knots are only slightly decelerated by
the medium in which they propagate (\citealt{bop07}). New ejected blobs
therefore travel into a medium constituted by strongly decelerated
blobs in light jets and by several high-speed blobs in heavy jets. The
average relative velocity between two consecutive blobs therefore
is expected to be higher in light jets than in heavy jets, leading in
general to X-ray emission in the former and to optical emission in the
latter case as a consequence of blob collisions.

At variance with our model predictions, however, all X-ray emitting
jets show also a knotty morphology in the optical band. A possible way
to reconcile model predictions and observations (i.e. reproducing both
X-ray and optical knots in a single run) is to consider the generation
of knots into a medium partly constituted by high-speed blobs and partly
by strongly decelerated blobs. In such a case, the relative velocity
between two consecutive blobs is expected to be small on average in the
part with high-speed blobs, leading to optical emission, and large in the
part with decelerated blobs, leading to X-ray emission. This scenario may
be reproduced by considering an ejection direction varying in time. If
the blob is ejected into a co-moving medium, filled of previously ejected
high-speed blobs, the average relative velocities between two
blobs is expected to be small, leading to optical knots. Otherwise,
if the blob is ejected into an almost stationary medium (because of the
strong deceleration of previously ejected blobs), the average relative
velocities between blobs is high and the knots resulting from the blob
interactions may be observed in the X-rays. The above scenario is
supported by the evidence that the location of the X-ray source in HH
154 is not completely aligned with the optical jet, as shown in Fig. 13 of 
\cite{bff08}.

\subsection{Interpretation of X-ray observations of HH jets}
\label{Main components of X-ray emission of the pulsed model}

As already discussed above, X-ray emitting jets are faint sources
and only limited spectral and morphological analysis can be performed
with the instruments in operation at the present time. The modeling of
the X-ray emission from protostellar jets can therefore be an important
tool in the interpretation of observations and may provide crucial
information to unveil the nature of the X-ray emission.

Our model has shown that a specific feature of a randomly pulsed jet
is the mutual interaction between plasma blobs that leads to irregular
patterns of knots aligned along the jet axis, possibly interacting with
each other. The knots emit preferentially in the X-ray band in light jets
and in the optical band in heavy jets. However,
although chains of optical knots are commonly observed in almost all HH
jets, current X-ray observations
do not allow to reveal the predicted knotty structure of X-ray jets due
to limited statistics. The only jet for which it was possible to analyze
the morphology of the X-ray source is HH 154, thanks to its proximity
and brightness (\citealt{fbm06}). In this case, a knotty structure in
the X-ray band has been revealed in agreement with our model predictions
and somewhat correlated to the observed knotty structure in the optical
band (\citealt{bff08}). Such a feature is probably common to all X-ray
emitting jets but is difficult to observe because of limited statistics.

\begin{figure}[!t]
\centerline{\psfig{figure=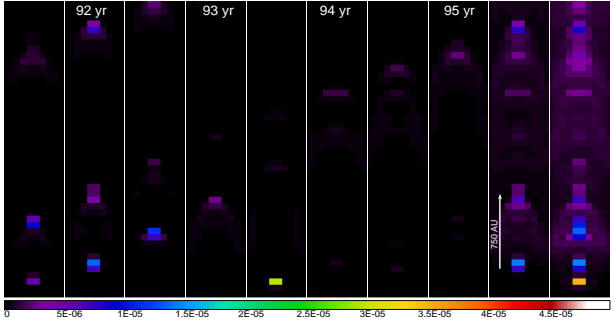,width=8cm}}
\caption{Evolution of the X-ray emission in the LJ0.5-M1000 run. First to
eighth panels: two consecutive panels are separated by a time interval of
$0.5$ yr, the first panel corresponding to $91.5$ yr since the beginning
of the jet/ambient interaction and the last to $95$ yr. The ninth panel shows
the $92$ yr frame, $93$ yr frame, $94$ yr frame, and $95$ yr frame added
together (the spatial scale is also shown: $750$ AU correspond to $5''$ at $150$ pc). The tenth panel shows the first to eighth frames added together.}
\label{X-05-somma}
\end{figure}

For instance, a knotty structure of the X-ray jet may be hidden
in the current observations of DG Tau. \citet{gsa08} analyzed the X-ray
source associated with this jet and needed to co-add several different
observations collected over a time-scale of 3 years to improve the
statistics. Consequently, any information on the spatial structure of
the X-ray source has been averaged over 3 years, making it impossible
to reveal its true morphology. In the co-added image, these authors
noted an elongated X-ray source with a spatial scale of about $5''$ and
originating from the central protostar. However, the observations of
DG Tau cannot exclude that the X-ray source is constituted by several
outgoing knots. Figure~\ref{X-05-somma} shows the evolution of the X-ray
emission for run LJ0.5-M1000 at the same resolution as the Chandra/ACIS-I
instrument (panels 1-8): once several different frames separated in
time by a few years have been added together (see last two panels in the
figure), the knotty morphology clearly recognized in the single frames
cannot be recovered in the summed image that shows an almost continuous
X-ray emitting structure along the jet axis, with spatial scales of
about $5''$, similar to that derived by \citet{gsa08}. In fact, even
if the proper motion of the single X-ray knot may be negligible in a
few years (as explained by \citealt{gsa08}), it is important to take
into account the fast variability of the structures within the jet and
the radiative time-scales, which determine the lifetime of the knots. A
knot can disappear over this time-scale and new ejected blobs can form
new observable knots.

Another important contribution to the X-ray emission from HH
jets comes from reverse shocks traveling in the opposite direction of
ejected blobs and possibly interacting with them (see Sect.~\ref{Spatial
distribution of X-ray emission}). These features are in general 
less frequent and short lived (lasting for less than about $5$ yr)
than the X-ray knots previously discussed, but are not negligible. They predict X-ray sources
within the jet with no appreciable proper motion and may explain, for
instance, the apparently stationary X-ray source detected at the base of
the optical jet HH\,154 over a time baseline of $4$ yr (\citealt{fbm06,
bff08}). An example of this feature is in Fig.~\ref{X-2-zoom} that
shows an X-ray emitting source roughly at the same position due to a
reverse shock powered by subsequent interactions with outgoing plasma
blobs. A series of observations of this source on a baseline of $4$ yr
may erroneously interpret the multiple interaction of the shock with
subsequent blobs as a stationary X-ray source located at the base of the
jet. This example shows how the comparison between model predictions
and observations may be a useful tool in the data
interpretation.

\subsection{Comparison between model predictions and observations of
X-ray emitting jets}
\label{Comparison between model predictions and observations of X-ray
emitting jets}

\begin{figure}[!t]
\centerline{\psfig{figure=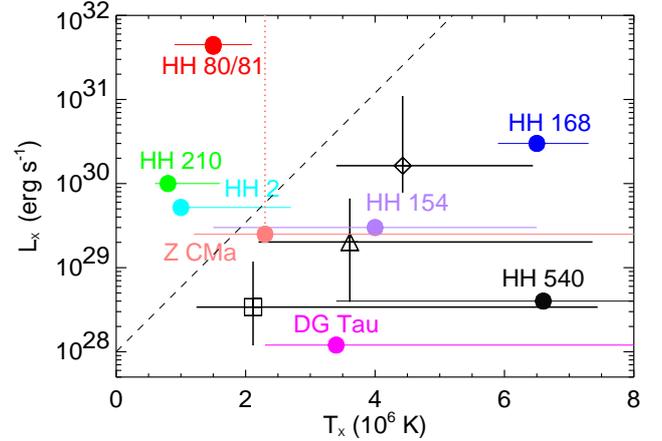,width=9.cm}}
\caption{X-ray luminosity as a function of best-fit temperature for the
three cases of ejection rate considered: $\Delta t = 0.5$ (diamond), $2$
(triangle), and $8$ (square) yr. The dots superimposed to the figure
mark the values derived from the observations of all known X-ray emitting
HH jets. The error bars of observed values are given if reported in the
literature. The X-ray luminosity of Z CMa is not well constrained and only
a lower limit is reported (pink dotted vertical line).}
\label{quantil}
\end{figure}

X-ray observations of HH jets allow us to derive the luminosity
and best-fit temperature of the X-ray sources associated with the jets
(see Table~\ref{tab-HH-X}). The comparison between these observed values
and those derived from our model may contribute with useful information on the properties of X-ray emission from protostellar jets. Figure
\ref{quantil} shows the X-ray luminosity, $L_{X}$, versus the best-fit
temperature, $T_{X}$, as a function of the ejection rate,
derived from the analysis of spectra synthesized from runs LJ0.5-M1000,
LJ2-M1000, and LJ8-M1000 (assuming high statistics, $10^{4}$ total
counts; see details in \citealt{bop07}). The symbols and the bars in
the figure represent the median values of $L_{X}$ and $T_{X}$
and the $10$th and $90$th percentiles ranges, respectively. Each
simulation predicts a wide range of variation for both the $L_{X}$
and the $T_{X}$ possibly related to the complex knotty
morphology and fast variability (a few years) of X-ray jets. As
expected, the most energetic case considered ($\Delta
t = 0.5 $ yr) leads to the highest luminosity and temperature:
the higher $\Delta t$, the lower $L_{X}$ and $T_{X}$.

Figure~\ref{quantil} also shows the X-ray luminosity and
temperature derived from X-ray observations for all the cases in
which some indication of $L_{X}$ and $T_{X}$ has been
reported in the literature (see Table~\ref{tab-HH-X}). Note that error
bars are given in Fig.~\ref{quantil} if reported in the literature. The
dashed line superimposed to the figure is arbitrary and separates HH
objects characterized by X-ray emission close to the base of the jet
(within $\approx 2000$ AU from the stellar source; to the right) from
HH objects showing X-ray emission at large distances (several thousands
of AU) from the young stellar object (YSO) from which the jet originates
(to the left; see Table \ref{tab-HH-X}). Figure~\ref{quantil} shows that
HH jets with X-ray emission at the base of the jet have luminosities and
temperatures in nice agreement with our model predictions. In particular,
the HH 154 jet is well described by a pulsed jet with an ejection rate
of two years. Remarkably, this result is in good agreement with the
independent estimate of morphological evolution time-scale of the optical knots discussed by \citet{bff08}. 
DG Tau is the weakest X-ray emitting jet (see Table
\ref{tab-HH-X}) and is consistent with the case of a jet with low ejection
rate (i.e. the case that shows the lowest luminosities). Conversely,
HH jets with X-ray emission at large distances from the YSO (namely
HH 80/81, HH 210, and HH 2) are characterized by very low values of
temperature ($\approx 1$ MK) and high luminosities ($> 10^{29}$ erg
s$^{-1}$) and do not match with our model results. The jet associated
with Z CMa appears to be intermediate to the two cases discussed above, the distance of the X-ray source from the YSO being $>2000$ AU.
In the framework of our model, we can interpret the X-ray knots located
further away from the YSO as due to strongly decelerated (and radiatively cooled) plasma blobs at low temperatures. In general, our model provides the
first attempt to describe the characteristics of almost all the X-ray emitting HH jets detected so far.

\subsection{Location of X-ray sources within the jet}
\label{Location of the X-ray source}

\begin{figure}[!t]
\centerline{\psfig{figure=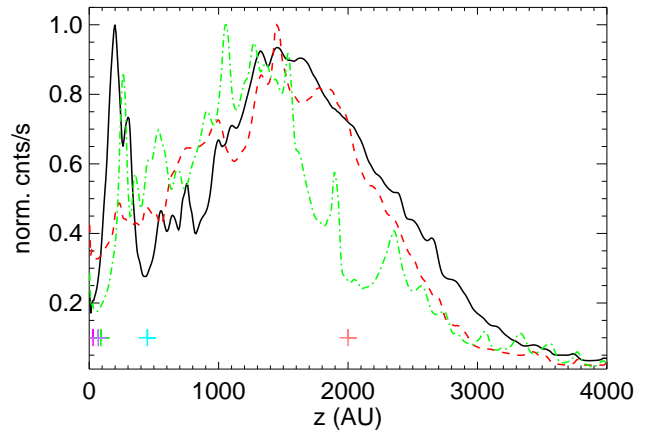,width=9cm}}
\caption{Normalized count rate in the band [0.3 - 10] keV integrated over the radial
direction and in time along the axis of the jet within $4000$ AU for the LJ0.5-M1000 (solid line in black), the LJ2-M1000 (dashed line in red), and the LJ8-M1000 (dashed-dotted line in green) run
after the computational domain has been fully perturbed by
the first blob ejected (time interval = from $40$ yr to $100$
yr for LJ0.5-M1000, from $\approx90$ yr to $\approx140$ yr for LJ2-M1000, and from $220$ yr to $400$ yr for LJ8-M1000).
The crosses superimposed indicate the position of observed X-ray
sources associated with HH jets for those cases that show emission close
to the base of the jet. From the nearest to the farthest from
the stellar source: DG Tau (in magenta), HH 154
(in violet), HH 540 (in green), TKH 8 (in cyan), and Z CMa (in red).}
\label{deltay-X}
\end{figure}

Another important parameter that allows us to compare model
predictions with observations is the position of X-ray sources within the
jet. In order to infer the position that is the most likely to
detect within the jet, we integrate the spatial distribution
of X-ray emission derived from the model both along the radial direction
and in time. This analysis requires that the computational domain has
been already fully perturbed by the first high-speed blob (ejected with
the maximum initial velocity) to avoid the initial transient phase. The
time interval of the integration therefore varies for the different
ejection rates considered: from $40$ to $100$ yr for $\Delta t = 0.5$ yr, from $\approx90$ yr to $\approx140$ yr for LJ2-M1000, and from $220$ to $400$
yr for $\Delta t = 8$ yr, the head of the jet traveling outside the
domain after $\approx 40$ yr in the former case and $\approx 220$
yr in the latter case. Figure \ref{deltay-X} shows the normalized count rate integrated along the jet axis and in time for the LJ0.5-M1000 (solid line in black), the LJ2-M1000 (dashed line in red), and the LJ8-M1000 run (dashed-dotted line in green). Most of the
emission is located at the base of the jet, within about $1500$ AU,
with a first bump within about $200$ AU from the YSO. The position of observed X-ray sources associated with
HH jets is superimposed to the figure for those cases that show emission
close to the base of the jet (i.e. HH jets to the right of the dashed
line in Fig.~\ref{quantil}). From the nearest to the farthest from the
protostar: DG Tau (in magenta), HH 154 (in violet), HH 540 (in green), TKH 8 (in cyan), and Z CMa (in red).

The X-ray emission is the largest (and, therefore, more easily
detectable) at the base of the jet. In fact, the interactions
among blobs and, possibly, among knots involve larger energies
close to the jet base than at larger distances and the outgoing knots
progressively fade as they propagate away from the stellar source
due to radiative cooling. We note that the position of observed X-ray
emitting sources associated with HH jets seems to be related to the
mass of the central object from which the jet originates. In particular,
almost all the LM objects show emission localized preferentially at the
base of the jet (see Fig. \ref{deltay-X}), with the
exception of HH 2. On the contrary, the X-ray sources detected in the
HM jets HH 80/81 and HH 210 are located at distances $>5\times10^{4}$
AU from stellar source (not reported in the figure).

\subsection{Predictions on future X-ray observations of HH 154}
\label{Predictions on the X-ray morphology of HH 154}

Among the known X-ray emitting jets, HH\,154 is the only one
that allows us to challenge the predictions of our model because of
its proximity and brightness. In particular, this is the only
object for which it was possible to analyze its morphology and the time
variability in the X-ray band to date (\citealt{fbm06}). The
HH 154 X-ray source has been detected as an almost point-like source in
2001 (\citealt{bfr03}) and as a knotty source, consisting of a stationary
(over a time baseline of $4$ yr) and an elongated source (showing a
proper motion of $\approx500$ km s$^{-1}$) in 2005 (\citealt{fbm06}). New future
observations of this object promise therefore to add important pieces
of information to shed light on the nature of the X-ray emission from
jets. Here we use the results of our pulsed jet model to
interpret future observations of HH\,154.

\begin{itemize}

\item[$\diamond\,\,$]{\textbf{Stationary source.}}
As discussed in Sect.~\ref{Main components of X-ray emission of
the pulsed model}, a pulsed jet with a random ejection speed can lead
to the same configuration (i.e. an apparent stationary source) if the
X-ray emission detected at different epochs arises from different
knots brightening roughly at the same position. However, the
probability to see an almost stationary source is lower than the probability to see single moving knots according
to our model, although not rejectable. In this case, therefore,
we cannot exclude alternative scenarios to that of a pulsed jet that
may explain the stationary source, as suggested by \cite{bfr03}. Among
these, the X-rays produced by the central protostar may be reflected
into our line of sight by a scattering layer located about 100-200 AU
from the parent star (a mechanism similar to that observed in Seyfert
2 galaxies). Alternatively, near the location of jet collimation the
dense medium or the magnetic field could act like a nozzle leading to
quasi-stationary shocks emitting X-rays (see discussions in \citealt{bop07}).

\item[$\diamond\,\,$]{\textbf{Previously detected sources show detectable
proper motion.}}
The new observations may show that the X-ray sources are
located in different positions with respect to previous observations. This
may happen because of the proper motion of the knots (not detectable
on time-scales of 4 years but measurable on a time-scales of 10 years)
or because new knots have emerged, while those observed in 2001 and 2005
have faded down, giving the impression of a motion of the sources.
Both these cases are predicted by our model.

\item[$\diamond\,\,$]{\textbf{New sources appeared.}}
Our model predicts that the brightness and the morphology of the
X-ray source associated with the pulsed jet can strongly change over
time-scales of the order of 10 years. We expect, therefore, that new
observations may show a very different morphology of HH\,154, with new
emerging knots and X-ray features (e.g. stationary shocks) catching the
previously existing ones. This is probably the most frequent configuration
predicted by the model.

\item[$\diamond\,\,$]{\textbf{Previously detected sources disappeared.}}
Since the last observation of HH 154 with Chandra has been performed
in 2005, the time-lapse between this and a new observation will
be at least of $5$ years, with a total time baseline since the
first observation in 2001 of about $10$ years. Over this time-scale,
the sources can both disappear, as also predicted by our model.
In fact, if the X-ray source is not powered by
new energetic blobs, its X-ray luminosity drops with a characteristic
time-scale of a few years.

\end{itemize}

\section{Conclusions}
\label{Conclusions}

We investigated the X-ray emission expected to arise from a randomly
pulsed jet with the aim to explain the nature of X-ray emission detected
in protostellar jets. We also explored the observable X-ray features
predicted to form as a consequence of the collisions between blobs
and knots within the jet, by exploring the parameter space given by
the ejection rate, initial jet Mach number, and initial density
contrast between the ambient medium and the jet. Our findings
lead to the following conclusions:

\begin{itemize}

\item[$\diamond\,\,$]{In all the cases, the interactions of the
ejected plasma blobs with the surrounding medium produce X-ray emitting
features. The main components contributing to the X-ray emission are:
irregular chains of knots; isolated high speed knots; steady knots;
reverse shocks; oblique shocks.}

\item[$\diamond\,\,$]{Light jets produce
significant X-ray emission consistent with the levels observed, whereas
heavy jets are characterized by very faint or no X-ray emission and emit
mostly in the optical band.}

\item[$\diamond\,\,$]{In the case of light jets (leading to
detectable X-ray emission), the X-ray luminosity is mainly determined
by the ejection rate of plasma blobs, rather than by the jet Mach number:
higher ejection rates are related to more energetic objects, thus leading
to higher X-ray luminosities.}

\item[$\diamond\,\,$]{Our model predicts X-ray luminosity and best-fit temperature in nice agreement with most of observed X-ray emitting jets. In particular the HH 154 jet is well described as a pulsed
jet with ejection rate $\Delta t = 2$ yr. Our model represents the first
attempt to describe all the X-ray emitting jets detected so far.}

\item[$\diamond\,\,$]{We found that most of the X-ray emission is
located at the base of the optical jet where the plasma blob collisions
are the most energetic and where, therefore, the probability to detect
X-ray emission is the highest. This result is consistent with the evidence
that almost all the LM objects show emission localized preferentially
at the base of the jet.}

\end{itemize}

In conclusion, our model explains why only a small fraction of HH jets has been detected in X-rays.
Indeed detectable X-ray emission may arise only under favorite conditions, namely high ejection rates of plasma blobs and high density contrast between the perturbed ambient medium and the ejected blob.

We stress that erroneous interpretations of observations can
easily be drawn if the X-ray sources (reverse shocks, stationary knots,
etc.) are not monitored frequently, since they show fast variability
over time-scales of a few years. On this respect, numerical models can
provide strong diagnostic tools to interpret observations. To this end,
we provided here model predictions on the morphology and characteristics
of the X-ray source associated with HH 154 that may be revealed in future
observations.

\begin{acknowledgements}

We would like to thank Joel Kastner and Ettore Flaccomio for providing us the spectral parameters of the COUP source HH 540.
R.B., S.O., M.M, and G.P. acknowledge support by the Marie Curie
Fellowship Contract No. MTKD-CT-2005-029768.  The software used in this
work was in part developed by the DOE-supported ASC/Alliances Center for
Astrophysical Thermonuclear Flashes at the University of Chicago, using
modules for thermal conduction and optically thin radiation constructed at
the Osservatorio Astronomico di Palermo. The calculations were performed
on the cluster at the SCAN (Sistema di Calcolo per l'Astrofisica
Numerica) facility of the INAF -- Osservatorio Astronomico di Palermo
and at CINECA (Bologna, Italy).  This work was partially supported by
Ministero Istruzione Universit\`{a} e Ricerca and by INAF.

\end{acknowledgements}

\bibliographystyle{aa}
\bibliography{references}

\end{document}